\documentclass[a4paper]{llncs}
\usepackage[T1]{fontenc}
\usepackage{hyperref}

\usepackage[utf8]{inputenc}

\usepackage{listings}
\usepackage{todonotes}

\usepackage{etoolbox}
\usepackage{xspace}
\newcommand{\curiesize}{\small}
\newcommand{\curieprefixes}{}
\newcommand{\defcurie}[2]{
  \expandafter\def\expandafter\curieprefixes\expandafter{\curieprefixes\texttt{#1}: & \begin{small}\url{#2}\end{small} \\[1ex] }
  \expandafter\newcommand\csname#1\endcsname[1]{{\curiesize \href{#2##1}{\nolinkurl{#1\ifblank{##1}{}{:##1}}}}}
}
\defcurie{dbo}{http://dbpedia.org/ontology/}
  
\defcurie{dbprop}{http://dbpedia.org/property/}
\defcurie{dbr}{http://dbpedia.org/resource/}
    
\defcurie{dcterms}{http://purl.org/dc/terms/}
\defcurie{foaf}{http://xmlns.com/foaf/0.1/}
\defcurie{gold}{http://purl.org/linguistics/gold/}
\defcurie{owl}{http://www.w3.org/2002/07/owl\#}
\defcurie{rdfs}{http://www.w3.org/2000/01/rdf-schema\#}
\defcurie{rdf}{http://www.w3.org/1999/02/22-rdf-syntax-ns\#}
  
\defcurie{schema}{http://schema.org/}
\defcurie{skos}{http://www.w3.org/2004/02/skos/core\#}
\defcurie{umbel}{http://umbel.org/umbel/rc/}
\defcurie{wiki}{https://en.wikipedia.org/wiki/}

\newcommand{\iri}[1]{<\url{#1}>}
\newcommand{\prefix}[2]{\href{#2}{\nolinkurl{#1}}: \url{#2}}
\newcommand{\iripart}[1]{\nolinkurl{#1}}
\newcommand{\sparql}[1]{{\curiesize \texttt{#1}}}

\newcommand{\demouri}{\url{http://purl.com/sparql-rest-api}}
\newcommand{\apipathpart}[1]{\begin{small}\nolinkurl{#1}\end{small}}
\newcommand{\apipath}[1]{\href{http://173.212.240.179:61001/api#1}{\apipathpart{#1}}}
\newcommand{\json}[1]{{\texttt{#1}}}
\newcommand{\rql}[1]{{\texttt{#1}}}

\let\origsubsubsection\subsubsection
\renewcommand\subsubsection{\vspace{-1em}\origsubsubsection}

\begin{document}

\title{Simplified SPARQL REST API}
\subtitle{CRUD on JSON Object Graphs via URI Paths}
\author{
Markus Schröder\inst{1} \and Jörn Hees\inst{1} \and Ansgar Bernardi\inst{1} \and\\
Daniel Ewert\inst{2} \and Peter Klotz\inst{2} \and Steffen Stadtmüller\inst{2}}

\institute{German Research Center for Artificial Intelligence GmbH (DFKI)\\ 
\email{\{markus.schroeder, joern.hees, ansgar.bernardi\}@dfki.de}
\and
Robert Bosch GmbH\\ 
\email{\{daniel.ewert, peter.klotz, steffen.stadtmueller\}@de.bosch.com} 
}
\maketitle

\begin{abstract}
Within the Semantic Web community, SPARQL is one of the predominant languages to query and update RDF knowledge.
However, the complexity of SPARQL, the underlying graph structure and various encodings are common sources of confusion for Semantic Web novices.

In this paper we present a general purpose approach to convert any given SPARQL endpoint into a simple to use REST API.
To lower the initial hurdle, we represent the underlying graph as an interlinked view of nested JSON objects that can be traversed by the API path.

\end{abstract}

\section{Introduction}
    Nowadays, the majority of developers already know how to use web technologies such as REST APIs and JSON.
    However, in order to use Semantic Web technologies, they typically still need extensive additional training.
    Before being able to perform simplistic CRUD (create, read, update, delete) workflows, they first need to learn about RDF basics, URIs, Literals, BNodes and how they're used to model knowledge as a graph of triples (as opposed to more often used JSON representations).
    Further, in the process newcomers are overwhelmed with a multitude of encodings, serialization and result formats, before finally being able to interact with a triple store via SPARQL Update (and to understand what they are doing).
    
    In this paper, we present an approach that aims to reduce this initial hurdle to use semantic technologies:
    We would like to allow Semantic Web newcomers to interact with a SPARQL endpoint without requiring them to go through extensive training first.
    To reach this goal, our approach transforms and simplifies a given SPARQL endpoint into a generic path based JSON REST API.
    During the design, we focused on simple CRUD workflows. 
    To reduce complexity, we decided against attempting to cover all SPARQL capabilities, but instead provide a trade-off between simplicity and expressivity.
    We use an easy to understand path metaphor to translate REST calls into corresponding SPARQL queries.
    Users can conveniently follow connections between the returned object views by iteratively extending the path of their requests.
    The resulting SPARQL response is translated back into an easy to understand and possibly nested JSON format.
    
    \begin{figure}[tb]
        \centering
        \includegraphics[width=\textwidth]{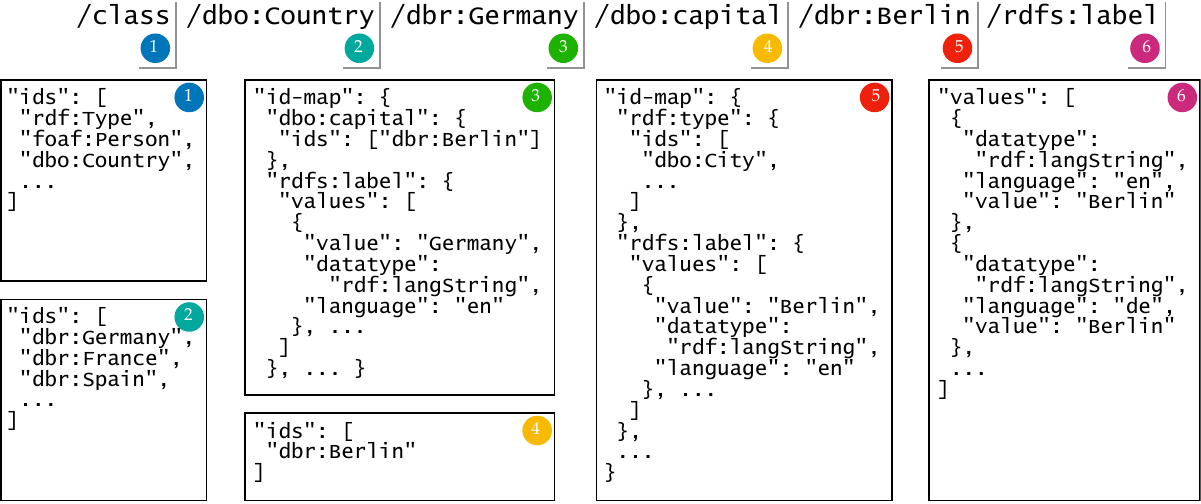}
        \caption{
            Illustration of basic REST API usage.
            Shown are excerpts of the JSON results for the different paths.
            Returned ids can conveniently be chained to walk the graph.
        }
        \label{fig:path_walk_example}
        \vspace{-1.2em}
    \end{figure}
    A simple example can be found in \autoref{fig:path_walk_example}.
    To get a list of all countries, users can access \apipath{/class/dbo:Country}.
    Chaining one of the returned ids, they can then access \apipath{/class/dbo:Country/dbr:Germany} and subsequently \apipath{/class/dbo:Country/dbr:Germany/dbo:capital} to arrive at \apipath{/class/dbo:Country/dbr:Germany/dbo:capital/dbr:Berlin} (or alternatively \apipath{/resource/dbr:Berlin}).

    An online demo can be found at \demouri .

\section{Related Work}\label{sec:relwork}
    Before detailing our approach, 
    we will briefly mention previous related works.
    
    Similarly to Battle and Benson \cite{Battle2008} our approach offers simple access based on class and resource entry points.
    However, our approach significantly extends the expressiveness by allowing arbitrary length paths to walk the graph and using wildcards and property paths.
    Further our serialization format consequently abstracts away from triples, as will be detailed in the following section.
    
    Anticipating SPARQL Update, Wilde and Hausenblas \cite{Wilde2009} discuss application of REST to SPARQL. 
    However, users would still have to master SPARQL and write RDF statements to communicate with the endpoint.
    
    The BASILar (Building Apis SImpLy) approach \cite{Daga2015} builds Web APIs on top of SPARQL endpoints by generating predefined REST resources.
    Similarly, grlc \cite{Merono-Penuela2016} builds Web APIs from SPARQL queries stored on GitHub.
    \cite{Garrote2011} describes a mapping approach from CRUD HTTP requests to SPARQL queries while
    \cite{Hopkinson2014} maps predefined REST calls to SPARQL queries, too.
    However, to work, these approaches need manual, up front definitions of mappings or SPARQL queries.

    Summarizing,
    in contrast to the mentioned works 
    our approach provides a zero-configuration REST to SPARQL conversion (evaluated during runtime) and uses well designed and easy understandable JSON objects (instead of triples).

\section{Approach}\label{sec:approach}

        As mentioned in the introduction, our approach aims to reduce complexity for Semantic Web newcomers by providing a simple to use, generic path based JSON REST API interface for a given SPARQL endpoint.
        To achieve this, we (bidirectionally) translate between the RDF predominant way of modelling knowledge in graph form and object oriented knowledge representations.
        The latter allows an easy to understand, nested serialization as JSON, which plays well together with our path walking metaphor.

    \subsubsection{Entry Points and Path Structure.}\label{sec:entrypoints}\label{paths}
        Focusing on common use-cases, our API interface provides two basic entry points: \apipath{/class} and \apipath{/resource}.
        The former is used to start with a known class browsing its instances, while the latter queries a resource by its known CURIE.
        An excerpt of the API's path grammar is shown in \autoref{lst:pathgrammar}.
        As can be seen, we allow arbitrary length traversal of the underlying graph by using the easy to understand folder metaphor:
        By alternating resources (RES) and properties (PROP) using the PATH rule we permit users to navigate the graph.
        
        \vspace{-1em}
        \begin{lstlisting}[caption={Excerpt of the API Path Grammar},label=lst:pathgrammar,gobble=12, xleftmargin=.1\textwidth, xrightmargin=.1\textwidth]
            API_PATH = "/api" (CLASS | RESOURCE)
            CLASS =    "/class" RES PATH RQL?
            RESOURCE = "/resource" PATH RQL?
            PATH =     ("/" RES "/" PROP)* ("/" RES)?
        \end{lstlisting}

    \subsubsection{Resulting JSON Syntax.}\label{sec:json_syntax}

        Our JSON objects are modeled along the outgoing edges of a node \sparql{x} (so all triples of the form \sparql{x ?p ?o}).
        We refer to nodes as ``ids'' using their CURIEs, to stress that users are not required to actually know that these are CURIEs or URIs (to them it is just an identifier).
        In simple use-cases the JSON object response contains one of two JSON array fields (\json{ids} or \json{values}) or a JSON object \json{id-map}, as can be seen in \autoref{fig:path_walk_example}.
        For consistency reasons, single id or value results are represented as arrays as well.
        
        Resources are listed in \json{ids} using their CURIEs.
        This reduces the need for escaping in paths, and generates more readable paths allowing easy manual entry.
        
        Literals are listed in \json{values}.
        To allow correct round-tripping, each of them is represented as object containing its value, language and datatype, inspired by the SPARQL JSON result representation.
        
        \json{id-map}s are used to map ids (resources or properties) to further components, for example to list a resource's properties.
        As we allow unbounded nested results, \json{id-map} naturally contain \json{ids}, \json{values}, \json{id-map} and \json{value-map} (see \autoref{sec:wildcards}).

    \subsubsection{HTTP CRUD Methods.}\label{sec:crud}

        Besides GET, our API supports POST, PUT and DELETE requests.
        The payload of POST and PUT is expected to be of the same structure as the GET results, allowing seamless round-trips. 
        In general, all modifying requests extend information from the body with that encoded in the path (e.g., \apipath{/class/X/x} implicitly adds the triple \sparql{x a X.}).
        POST requests generate (and return) a new random identifier for the object specified in the body, while PUT requests are used to update an object identified by the path.
        Depending on the path depth, DELETE requests delete a full resource (depth: 1, all in- and outgoing edges), the specified outgoing properties (depth: 2) or one specific triple (depth: 3).

    \subsubsection{Extended Expressiveness: Wildcards, Property Paths \& RQL.}\label{sec:wildcards}\label{sec:property_paths}\label{sec:rql}
        Apart from these basic features, we extended our API with a couple of noteworthy features that seamlessly integrate into the path metaphor.
        
\begin{figure}[tb]
            \centering
            \includegraphics[width=\textwidth]{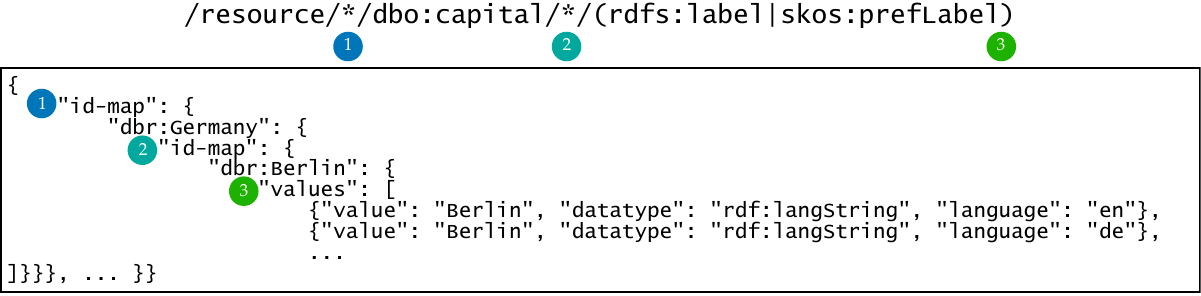}
            \caption{
                Example of path extensions using wildcards and a property path: The example lists all object ids with an outgoing \dbo{capital} edge, their corresponding linked object ids and their corresponding \rdfs{label} or \skos{prefLabel} as values.
                As can be seen each \texttt{*} introduces an extra nesting level, while property paths (as in SPARQL) are transparently collapsed in the result.
            }
            \label{fig:path_extensions}
            \vspace{-1em}
        \end{figure}

        A very powerful feature is the well known Bash wildcard \apipathpart{*}, which we allow in any RES and PROP position in the path.
        The asterisk is interpreted in an ``all of them'' way, introducing an additional nesting level for each asterisk in the resulting JSON.
        It allows to quickly create a partial view of nested objects, as can be seen in \autoref{fig:path_extensions}.

        As also shown in \autoref{fig:path_extensions}, we additionally permit the use of SPARQL property paths\footnote{\url{https://www.w3.org/TR/sparql11-query/\#propertypaths}} in every PROP position by using surrounding brackets (e.g. \apipathpart{/:x/(foaf:name|rdfs:label)/}).
        Reminding of regular expressions, this enables queries containing alternatives, inverse directions and multiple hops.
        Similarly to SPARQL, the followed property path is not shown in the result (collapsed).
        
        Combining wildcards and inverse property paths also allows us to step over literals:
        For example, \apipath{/resource/*/foaf:name/*/(^rdfs:label)} will list all object ids that link to values via \foaf{name}, the corresponding Literals, and (other) object ids which use the same literal as \rdfs{label}.
        As Literals are complex objects they cannot appear as keys in JSON syntax.
        Hence, we introduce a last additional keyword to our JSON result format:
        \json{value-map}, which represents mappings of values to further components as a list of pairs.
        
        Apart from wildcards and property paths, we allow the API paths to be extended with Resource Query Language (RQL)\footnote{\url{https://github.com/persvr/rql}} methods, such as \rql{regex}, \rql{sort}, \rql{limit} and aggregations like \rql{count}, \rql{sum} and \rql{avg}.

    \subsubsection{Further Features: Batch \& BNode Handling.}

    We additionally implemented batch processing in order to bundle many similar requests into one and to reduce connection overhead.
    To avoid URI length restrictions and because processing a batch usually is a procedure, we implement it via JSON-RPC\footnote{\url{http://www.jsonrpc.org/specification}}.
    Moreover, a \apipath{/namespace} entry point can be used to to resolve prefixes, e.g. \apipath{/namespace/rdfs,owl}.
    
    Our API handles BNodes via their fixed (skolem) URIs as mentioned in \cite{Mallea2011BNodes} and supported by many triplestores (e.g., CURIE \sparql{\_:b1} $\Leftrightarrow$ URI \sparql{<\_:b1>})).
    This allows users to use and traverse BNodes like normal URIs.

\section{Conclusion and Outlook}\label{sec:concl}
    In this paper we presented an approach to turn any given SPARQL endpoint into a simple to use JSON REST API.
    To achieve this, our approach translates between CRUD API requests and SPARQL (Update) queries.
    The API paths allow users to simply navigate the underlying graph to their point of interest.
    The paths further allows wildcards and SPARQL property path components, seamlessly integrated in the API as deeper nestings of the resulting JSON.
    
    While the development of our approach already embeds a lot of user feedback, in the future we would like to enhance our approach by adding further ideas.
    Apart from improvements in the areas of error messages and content negotiation, we especially would like to focus on path based permissions and easy to use CRUD operations for simplistic TBox management.
    
    A live demo showing various examples and their corresponding generated SPARQL queries, the source code, API docs and further information are available online:
    \demouri

\begin{tiny}
\bibliographystyle{llncs}
\bibliography{paper}
\end{tiny}

\nopagebreak

\end{document}